# AI-Designed Photonics Gratings with Experimental Verification

Yu Dian Lim[1,*], Chuan Seng Tan[1,2*]

[1]*School of Electrical and Electronics Engineering, Nanyang Technological University, 639798 Singapore*
[2]*Institute of Microelectronics, Agency for Science, Technology and Research (A*STAR), 117685 Singapore*
*\*yudian.lim@ntu.edu.sg, \*tancs@ntu.edu.sg*

**Abstract:** Artificial Intelligence (AI) software based on transformer model is developed to automatically design gratings for possible integrations in ion traps to perform optical addressing on ions. From the user-defined (x,z) coordinates and full-width half-maximum (FWHM) values, the AI software can automatically generate the Graphic Design System (GDS) layout of the grating that "shoots" light towards the pre-defined (x,z) coordinates with built-in finite-difference time-domain (FDTD) simulation for performance verification. Based on the FDTD verification, AI-design gratings produced grating-to-free-space light that "shoots" towards the provided (x,z) target with < 2 μm deviations. For most attempts, the FWHM of FDTD simulation has < 2 μm deviations from the user-defined FWHM. The AI-designed gratings were successfully taped out and capable of producing output light for possible optical addressing of trapped ions.

## 1. Introduction

Over the past decade, artificial intelligence (AI) has been widely adopted to solve complex designing issues in silicon photonics devices [1]. Adibnia *et al.* reported the inverse design of FBG-based optical filters using convolutional neural network (CNN) approaches [2]. Pal *et al.* reported the inverse design of micro-ring resonators using machine learning techniques [3]. In our research group, we have been using AI techniques in our photonics design, such as integration of trained models to assist the lengthy finite-difference time-domain (FDTD) simulation [4,5], recognizing both FDTD-simulated and experimentally-measured beam profiles [6–8], or auto-focusing of light beam using object detection model [9].

One of the use case scenarios of AI-assisted photonics device design is the design of grating couplers for possible integrations in ion traps. The concept of photonics integration involves burying grating couplers underneath the planar electrodes in ion trap to "aim" light from gratings to trapped ions for optical addressing required for quantum computing operations [10,11]. However, some of the key challenges from using this approach involve the design and placement of gratings in the ion trap so they would produce light beams that accurately "shoot" towards the trapped ions [12,13]. Besides the accuracy, the desired light beams should also have high degree of focus to enable precise optical addressing on each ion. At the same time, the optical addressing of trapped ions involves wide range of wavelengths. For instance, $^{88}Sr^+$ ions require optical addressing using laser wavelengths from 405 nm to 1,092 nm for quantum computing operation. Thus, developing a standard process design kit (PDK) specifically for the photonics integration of ion traps can be challenging [14].

Theoretically, the direction of the grating-to-free-space light ($\theta$) can be described with the following equation:

$$\sin\theta = \frac{n_{eff} - \frac{\lambda}{\Lambda}}{n_{clad}}$$

where $n_{eff}$, $n_{clad}$, $\lambda$, and $\Lambda$ are effective index of the guided mode in the waveguide, refractive index of the $SiO_2$ cladding, targeting wavelength, and grating pitch, respectively [15]. Thus, to enhance the degree of focus of the light beams coupled out from the buried gratings, we have previously reported a mixed pitch grating with 1.1μm/1.2μm pitches in the same grating coupler. These similar pitches produced light beams with higher degree of focus than their single-pitch counterparts [16]. To further enhance the flexibility of the mixed pitch gratings, the selection of pitches can be varied, not limiting to 1.1μm/1.2μm. At the same time, more pitches (such as 3 different pitches) can be involved, with additional gaps between each pitch region. However, these significantly increase the number of tunable parameters involved. At the same time, in designing mixed pitch gratings, different pitches not only associated with direction of the grating-to-free-space light ($\theta$) but also yield different light intensity. Thus, when designing the mixed pitch grating, we cannot just obtain the $\theta$ values of each pitch and compute the possible focal point of the grating by geometrical calculations.

To resolve this problem, artificial intelligence (AI) models, such as transformer model, can be used to assist the design of mixed pitch gratings. Transformer model was first introduced by Google Inc. mainly for its application in natural language processing (NLP) [17]. In our research group, we have previously reported several works that used transformer model to assist in the design of gratings for their integrations in ion traps [4,6,8,18]. In this work, we demonstrate the usage of AI system for the inverse design of 3-pitches gratings based on transformer model.

## 2. Methodology

Fig. 1(a) illustrates the optical addressing of trapped ion by 3-pitches grating. From Fig. 1(a), grating coupler is buried underneath the planar ion trap. In optical addressing, laser light of specified wavelengths (e.g. 405, 422, 674, 1033, 1092 nm for trapped $^{88}Sr^+$ ion) are propagated through the waveguide towards the mixed pitch grating. By varying the grating parameters, we can control the (x,z) coordinate of the grating-to-free-space light's focal point. The detailed architecture of the 3-pitches grating is shown in Fig. 1(b). As referred to Fig. 1(b), 'p' represents the pitch in μm (0.3 - 1.2 μm) and 'c' represents the number of protrusions ('count', 2 - 12) in region 1, 2, or 3. Two gaps, gap1 and gap2, range between 0 to 2 μm, are present.

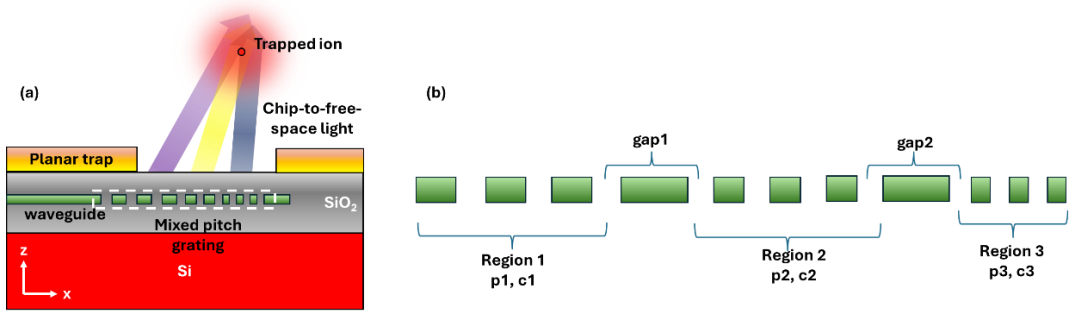

Fig. 1. (a) Illustration of optical addressing using 3-pitches grating, (b) Key parameters of 3-pitches grating

To demonstrate the usage of AI in designing 3-pitches gratings, we developed a full software architecture shown in Fig. 2. First, the user will define the (x, z) coordinate where the grating-to-free-space light should be focused on. Based on the user input, a mock electric field (E-field) curve will be generated using Gaussian equation. This is to tell the AI system that "a grating that can produce such E-field curve, peaked at given x-coordinate, located at given z-coordinate is needed". The mock E-field curve will be used as the input, while the trained transformer model will generate the associated grating parameters (ref. Fig. 1(b)) that meet the given requirement. The grating parameters will be subjected to finite-difference time-domain (FDTD) simulation for performance verification. Once the performance is verified (the generated grating parameters meet the (x,z) requirements specified by the user), Graphic Design System (GDS) layout of the grating will be generated. For experimental benchmarking, we fabricated the generated GDS layouts to perform experimental measurements on them. Under 'Experimental Benchmarking' in Fig. 2, the test structure consists of an input grating, a reference grating, and AI-designed gratings using the parameters suggested by the trained transformer model. The design methodology of the test structure is described in ref. [8]. Both FDTD simulations and experimental measurements in this work are carried out on 220 nm silicon-on-insulator (SOI) platform using under 1550 nm laser light. To obtain the beam profiles of the grating-to-free-space light, infrared (IR) camera (Ophir XC-130) is used.

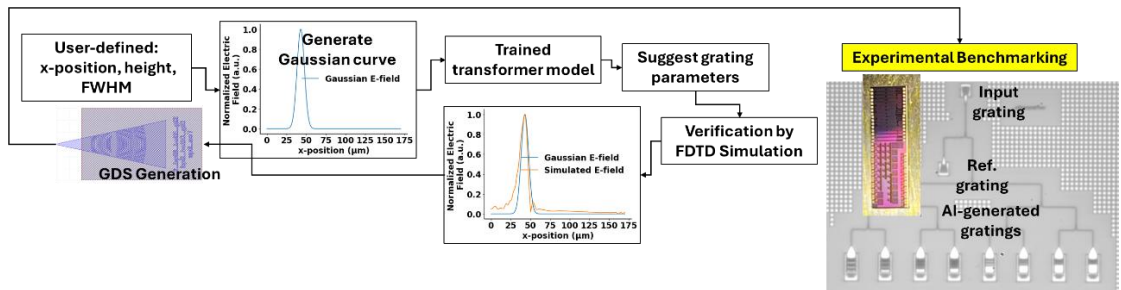

Fig. 2. Full software architecture developed for the AI-generated gratings

## 3. Model Training

For the preparation of "Trained transformer model" illustrated in Fig. 2, we prepared 10,000 different combinations of parameters in 3-pitches gratings (ref: Fig. 1(b)) for FDTD simulation to obtain the grating-to-free-space propagation of light. When training the model, the electric field (E-field) spectrum along x-axis and their respective z-position (axis ref.: Fig. 1(a)) are used as the input, while the 8 key parameters are used as the output. As referred to Fig. 3, two-dimensional (2D) E-field of grating-to-free-space light from gratings with 10,000 different combinations of parameters are simulated using FDTD technique. Under various heights (along z-axis), one-dimensional (1D) E-field spectrum are extracted. The spectrum and the associated heights are used as the input of the transformer model, while the 8 parameters shown in Fig. 1(b) is used as the output of the transformer model. The full Python code of the transformer model is given in ref. [19]. After the model training, the transformer model is deployed and used as the "Trained transformer model" shown in Fig. 2.

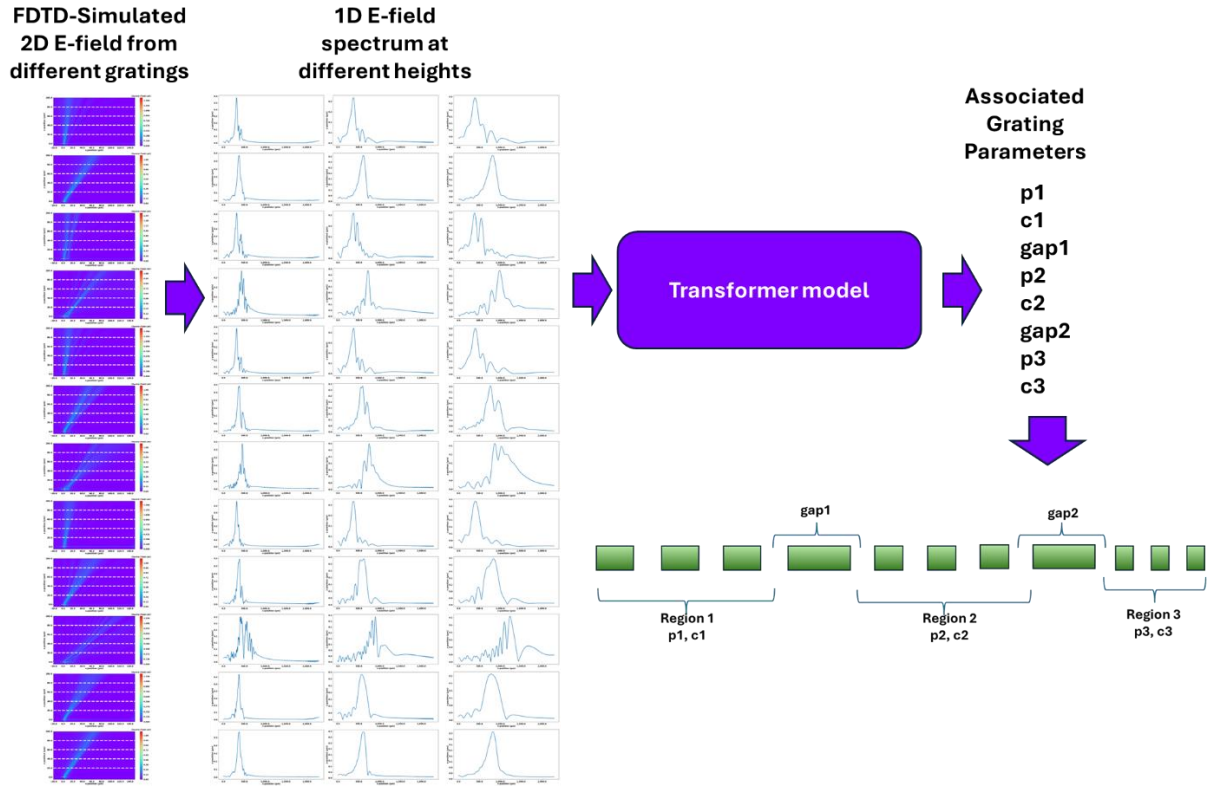

Fig. 3. Illustration of model training

## 4. Model Testing and Experimental Benchmarking

The use-case-scenario of Fig. 2 will be demonstrated in this section. As shown in Fig. 4, (x,z) coordinates of (22.2µm,30µm), (7.8µm,40µm), (34.5µm,50µm), and (57.9µm,90µm) are attempted. Going through the process illustrated in Fig. 2, the specified coordinates are first used to generate mock E-field spectrum using Gaussian equation. Then, the E-field spectrum will be fed into the trained transformer model for grating parameters suggestion. Based on the suggested parameters, FDTD simulation is performed to verify that the grating-to-free-space light can "hit" the given coordinates Fig. 4(a, e, i, m). At the same time, the 1D E-field spectrum at the given height are extracted and compared against the mock E-field spectrum generated by Gaussian equation Fig. 4(b, f, j, n). If the FDTD-simulated E-field spectrum matched the Gaussian-generated mock E-field spectrum, the GDS layout of the gratings with be generated, as shown in the inset of Fig. 4(c, g, k, o). The GDS layouts are then used to construct test structures shown in the 'Experimental Benchmarking' part of Fig. 2. The test structures are then fabricated and tested using an IR camera. The beam profiles are shown in Fig. 4(c, g, k, o). Then, the field intensities along x-axis are extracted and compare against the Gaussian-generated mock E-field and FDTD-simulated E-field. For better comparison, the E-field (or intensity) values and x-positions are normalized for all spectrum, as shown in Fig. 4(d, h, l, p).

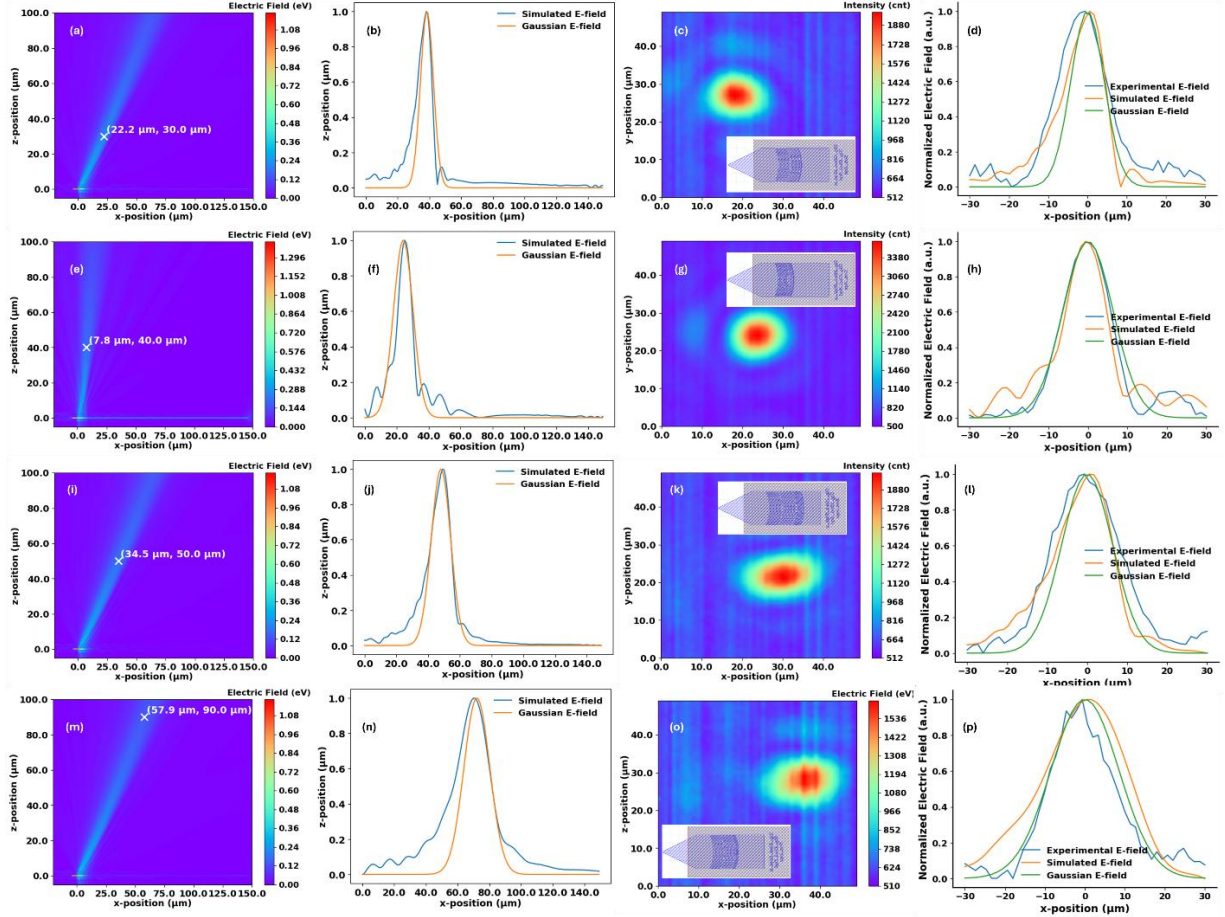

Fig. 4. (a, e, i, m): FDTD-simulated 2D E-field distributions, (b, f, j, n): FDTD-simulated 2D E-field distributions, (c, g, k, o): measured beam profiles using IR camera (inset: GDS layout of the grating), (d, h, l, l): Normalized Gaussian-generated, FDTD-simulated, and experimentally-measured E-field spectrum

Based on Fig. 4(d, h, l, p). the comparisons between Gaussian-generated, FDTD-simulated, and experimentally measured spectrum are shown in Table 1. Generally, the positions of the peak E-field for both Gaussian-generated and FDTD-simulated spectrum are similar with < 2 μm deviations. This implies that the grating parameters suggested by the trained transformer model can produce grating-to-free-space light beams that are able to "hit" the user-defined (x,z) coordinates. However, as the IR camera used in this work has relatively low pixel count (~50×50 pixels in a typical light beam), experimental measurement of light propagation from grating to free space can be challenging.

Comparisons between the full-width half-maximum (FWHM) of all three spectrum (Gaussian-generated, FDTD-simulated, and experimentally measured) can be observed in Fig. 4(d, h, l, p). and Table 1. For most (x,z) coordinates, the FWHM of Gaussian-generated and FDTD-simulated E-field has <2 μm deviations. The relatively large deviation (~3.5 μm) present in (7.8μm,40μm) coordinate could be attributed the absence of training data, as the angle of propagation of such (x,z) coordinate is unusually small (~11° from the vertical) as compared to other coordinates presented in Fig. 4(a), (i), and (m). Comparing the FWHM between simulated and experimental E-field, the deviations are typically < 5 μm. The sources of deviations can be attributed to several factors, including manufacturing flaws and low pixel count of the IR camera. Nevertheless, note that the deviations between FWHM values of FDTD-simulated and experimentally measured E-field are not directly related to the accuracy of the trained transformer model.

On top of the results presented in Fig. 4 and Table 1, we have also prepared a video demonstration on the whole process presented in Fig. 2. The video demonstration that generates GDS layouts of other possible coordinates (ref. [20]).

Table 1: Comparison between Gaussian E-field, Simulated E-field, and Measured Intensity

| (x,z) coordinates | | Gaussian E-field | Simulated E-field | Measured Intensity |
|---|---|---|---|---|
| (22.2μm,30μm) | Peak E-field along x-axis | 22.2 μm | 22.2 μm | - |
| | FWHM | 10μm | 11.33μm | 14.36μm |
| (7.8μm,40μm) | Peak E-field along x-axis | 7.8 μm | 6.7 μm | - |
| | FWHM | 15 μm | 11.57 μm | 14.14 μm |
| (34.5μm,50μm) | Peak E-field along x-axis | 34.5 μm | 33.33 μm | - |
| | FWHM | 15 μm | 15.5 μm | 19.14 μm |
| (57.9μm,90μm) | Peak E-field along x-axis | 57.9 μm | 59.99μm | - |
| | FWHM | 20 μm | 20.9 μm | 18 μm |

## 5. Summary

In this work, we have developed an AI software architecture that automatically designs the grating layouts for their integration in ion traps based on the user-defined (x,z) coordinates and FWHM values. Four (x,z) coordinates are attempted and reported. It was found that the AI-designed gratings produced grating-to-free-space light that hits the user-defined coordinates with < 2 μm deviations along x-axis. For the FWHM, most AI-designed gratings produced light beams with FWHM of < 2 μm deviations from the user-defined FWHM. The FWHM deviation value of (7.8μm,40μm) coordinate is relatively larger than the other attempted coordinates (~3.5 μm) due to the scarcity of relevant training data. The AI-designed gratings have been taped out successfully and capable of producing output light for possible optical addressing of trapped ions. The FWHM deviations between experimentally measured and FDTD-simulated beams are typically < 5 μm.

This work was supported by Ministry of Education Tier-1 RG135/23, RG67/25, and RT3/23. The taped out in this work was done in Advanced Micro Foundry (AMF) under its multi-project wafer (MPW) services for passive devices.